\documentclass[10pt]{article}
\usepackage{graphicx,fancyhdr,amssymb}

\begin{document}
\begin{center}
{\large\bf A Model for Counterparty Risk with Geometric  Attenuation Effect\\ and the Valuation of CDS}\\
\bigskip
 Yunfen BAI $^{\dag\ddag}$ ,\,\, Xinhua HU$^\dag$\ \ Zhongxing
YE$^\dag$
\end{center}
\par
\centerline{\small{$\dag$  Department of Mathematics,\ Shanghai
Jiaotong University,\ Shanghai 200240, China}} \centerline{\small
$\ddag$ Department of Mathematics,\ Shijiazhuang College,\
Shijiazhuang, Hebei 050035, China\ }
\renewcommand{\thefootnote}\*\footnote{$*$Project supported by the National
Natural Science Foundation of China (No.70671069).}
\renewcommand{\thefootnote}\*\footnote{Corresponding
authors Yunfen BAI, E-mail: baiyun200588@126.com; Zhongxing Ye,
E-mail: zxye@sjtu.edu.cn}

\centerline{January, 2007}
\bigskip

\begin{quote}
{\bf Abstract:}\,\,\,{\small In this paper, a geometric  function is
introduced to reflect the attenuation speed of impact of one firm's
default to its partner. If two firms are competitions (copartners),
 the default intensity of
one firm will decrease (increase) abruptly when the other firm
defaults. As time goes on, the impact will decrease gradually
until extinct. In this model, the joint distribution and marginal
distributions of default times are derived by employing the change
of measure,
 so can we value the fair swap premium of a CDS.}\\
{\bf\small Key words:}\,\,{\small Dependent default;\,\,Geometric
attenuation function;\ \ Change of measure;\ \
 Credit Default Swap(CDS) }
\end{quote}
\par{\bf 2000 Mathematics Subject Classification: 62P05}\\
%\par {\bf  Chinese Library Classification(GB/T13745):\,\, F830}\\
\par
\bigskip
{\bf{1. Introduction}}
\par
The rapid expansion in recent years of market for the credit
derivatives had led to a growing interest in the valuation of these
instruments including the credit default swaps(CDS).The reference
issuers and the  derivative issuers not only have default risk, but
also correlate in some way.  As remarked by Jarrow and Yu (2001),"an
investigation of counterparty risk is incomplete without studying
its impact on the pricing of credit derivatives". We can distinguish
three different approaches to model default correlation in the
literature of intensity credit risk modeling. The first approach to
model default correlation makes use of copula functions. A copula is
a function that links univariate marginal distributions to the joint
multivariate distribution with auxiliary correlating variables. Li
(2000) was probably the first to explicitly use the concept of
copulas in the context of basket default derivatives pricing.
\par
The second approach introduces correlation in  firms' default
intensities making them dependent on a set of common variables $X_t$
and on a firm specific factor. These models have received the name
of conditionally independent defaults (CID) models, because
conditioned to the realization of the state variables $X_t$ the
firm's default intensities are independent as are the default times
that they generate. for example Duffie and Singleton (1999) and
Lando (1994, 1998).
\par
The last approach to model default correlation, contagion models,
relies on the works by Davis and Lo (1999) and Jarrow and Yu (2001).
It is based on the idea of default contagion in which, when a firm
defaults, the default intensities of related firms jump upwards.
%In these models default dependencies arise from direct linkage
%between firms. Default of one firm increases the default
%probabilities of related firms, which might even trigger the
%default of some of them.
 Leung and Kwok (2005) gave the analytic
solution for the CDS premium for Jarrow and Yu (2001)model by
employing the change of measure which was introduced in
Collin-Dufresne (2004). But it is unrealistic for them to assume
that one firm's default intensity keep a constant jump  after the
other firm defaults. In this paper, we introduce a geometric
 function to reflect the attenuation impact of one firm's default
to other firms' default intensities. In our model, one firm's
default will influence other firms' default intensities and the
impact will decrease until extinct as time goes on. That is to say
after a period of time, one firm's default intensity will depend
the firm itself while the impact of other firm's default will
disappear. The model is more realistic.
\par
The paper is organized as follows. In  section 2, we introduce a
 geometric   function  to reflect the
attenuation impact of  one  firm's default to its partners, and give
emphasis on the case when the two firms are competitions. The joint
and marginal distributions of the two firms' default times are got
by employing the change of measure which was introduced in
Collin-Dufresne (2004). In section 3, we price the CDS premium using
the conclusion of section 2 and get the analytic solution. The paper
is ended with conclusion in the last section.
\par\bigskip
{\centerline\bf\large{2.  A Model for Dependent Default with
Geometric  Attenuation function}}
\par\bigskip
We consider an uncertain economy with a time horizon of $T^*$
described by a filtered space $(\Omega,\mathcal{F},\
\{\mathcal{F}_t\}_{t=0}^{T^*}, \ \mathcal\bf{P})$ satisfying
$\mathcal{F}=\mathcal{F}_{T^*}$, where $\mathbf{P}$ is the
risk-neutral(equivalent martingale) measure in the sense of Harison
and Kreps (1981), that is, all security prices discounted by the
risk-free interest rate $r_t$ are martingale under $\mathbf{P}$.
\par
In this section, we construct a two-firm model with default
correlation. Suppose firm B and firm C have high direct linkage
which are competitions or copartners. The default time of firm i
(i=B,C) is denoted by $\tau^i$ which posses a strictly positive
$\mathcal{F}_t$-predictable intensity process $\lambda_t^i$ with
right-continuous sample paths. If we define $N_t^i=I_{(\tau^i\leq
t)}$ as the default indicator function which equals to 0 before
%Õâ¾ä»°ÓÐÎÊÌâ
 firm i defaults and 1 otherwise.
 \par
Let
 $\mathcal{H}_t=\sigma(N^B_{s},\ 0\leq s\leq t)\vee\sigma(N^C_{s},\ 0\leq s\leq t)$, then
we can get that
$$M_t^i=N_t^i-\int^{t\Lambda \tau^i}_0 \lambda_s^i ds\eqno(1)$$
is an $\mathcal{F}_t$-local martingale and the conditional
survival probability of firm i is given by
$$P(\tau^i>T|\mathcal{H}_t\ \vee\ \mathcal{F}_t)=I_{(\tau^i> t)}
E[\exp(-\int^T_t \lambda^i_s ds)|\mathcal{F}_t].\eqno(2)$$  The
default correlation between firm B and firm C is characterized by
the correlated default intensities:
$$\lambda_t^B=b_0+I_{(\tau^C\leq t)}\frac{b_1}{b_2(t-\tau^C)+1},\eqno{(3)}$$
$$\lambda_t^C=c_0+I_{(\tau^B\leq t)}\frac{c_1}{c_2(t-\tau^B)+1},\eqno{(4)}$$
where $b_0,\ c_0,\ b_2,\ c_2$ are nonnegative real numbers, and
$b_1,c_1$ are real numbers satisfying $b_0+b_1>0,\ c_0+c_1>0$. In
this model, the default of firm C(B) will bring abrupt change to the
default intensity of firm B(C). If B is one competition (copartner)
of firm C, $b_1<0(>0)$, and when firm C defaults, the default
intensity $\lambda_t^B$ jumps by the amount of $|b_1|$ from $b_0$ to
$b_0+{b_1}$. As time goes on, the impulsion effect will attenuate
until extinct with geometric speed, that is to say, $\lambda_t^B$
will come back to $b_0$ at last. This explanation is the same as C
is one competition(copartner) of firm B. Parameters $b_0$ and $c_0$
are the firm-specific default intensity. Parameters $b_1$ and $ c_1$
reflect the impact intensities of counterparty's default, and when
$b_1=0$ and $c_1=0$, firm B and firm C are default-independent.
Parameters $b_2$ and $c_2$ are non-negative real numbers, which
reflect the attenuation speed. When $b_2=0$ and $c_2=0$, the model
becomes the one in Jarrow and Yu (2001) and Yuen and Yue (2005).
\par\bigskip
To calculate the joint distribution of the two default time of firm
B and firm C in $[0,T]$($T<T^*)$, we adopt the change of measure
introduced by Collin-Dufresne (2004), defining a firm-specific
probability measure $P^i$ which puts zero probability on the pathes
where default occurs prior to the maturity T. Specifically, the
change of measure is defined by
$$Z^i_T\doteq \frac{dP^i}{dP}{\bigg{|}}_{\sigma(\mathcal{H}_T\vee\mathcal{F}_T)}=
I_{(\tau^i> T)}\exp(\int^T_0 \lambda^i_s ds),\eqno(5)$$
 where $P^i$ is a firm-specific(firm i) probability
measure which is absolutely continuous with respect to $P$ on the
stochastic interval $[\tau^i,+\infty)$. To proceed the calculation
under the measure $P^i$, we enlarge the filtration to
$\mathcal{G}^i=(\mathcal{G}_t^i)_{t\geq 0}$ as the completion of
$\sigma(\mathcal{H}_t\vee\mathcal{F}_t)_{t\geq 0}$ by the null set
of the probability measure $P^i$. One can show that $Z_T^i$ is a
uniformly integrable P-martingale with respect to $\mathcal{G}_T^i$
and  almost surely strictly positive on $[0,\tau^i)$ and almost
surely strictly equals to zero on $[\tau^i,+\infty)$(see
Collin-Dufresne (2004)).
\par
Under the default risk structure specified in Eqs.(3) and (4), the
survival probabilities of firm B and firm C are defined
recursively through each other and this leads to the phenomenon of
"looping default". Under the new measure $P^B$ defined by Eq.(5),
$\lambda_t^C=c_0$ for $t<\tau^C$, this effectively neglect the
impact of firm B's default on the intensity of firm C, and looping
default no longer exists. An analogous argument also holds under
the measure $P^C$.
\par\bigskip
{\bf Proposition 1}\ \ When $-b_1=b_2=b>0$ and $-c_1=c_2=c>0$, the
joint distribution of default times $(\tau^B,\tau^C)$ with the
default intensities defined by Eqs.(3) and(4) on $[0,T]\times
[0,T]$ is found to be
$$ P(\tau^B>t_1,\tau^C>t_2)=\left\{
\begin{array}{ll}
c(t_2-t_1+\frac{1}{c}-\frac{1}{b_0}){e^{-b_0t_1-c_0t_2}}+\frac{c}{b_0}e^{-(b_0+c_0)t_2},&
\ \ \mbox{for}\  t_1\leq t_2\leq T \\
b(t_1-t_2+\frac{1}{b}-\frac{1}{c_0}){e^{-b_0t_1-c_0t_2}}+\frac{b}{c_0}e^{-(b_0+c_0)t_1},&
\ \ \mbox{for}\ t_2<t_1\leq T ,
\end{array}
\right.\eqno(6)
$$
and the joint density is
$$ f(t_1,t_2)=\left\{
\begin{array}{ll}
c{b_0}{c_0}[(t_2-t_1)+\frac{1}{c}-\frac{1}{c_0}]e^{-b_0t_1-c_0t_2},&\ \ \mbox{for}\  t_1\leq t_2\leq T \\
b{b_0}{c_0}[(t_1-t_2)+\frac{1}{b}-\frac{1}{b_0}]e^{-b_0t_1-c_0t_2},&
\ \ \mbox{for}\ t_2<t_1\leq T .
\end{array}
\right.\eqno(7)
$$
\par\newpage
{\bf Proof:}\par
 If $t_2\leq t_1\leq T $,
 $$(\tau^B>t_1,\tau^C>t_2)\in
\mathcal{H}_{t_1}\subset\mathcal{G}_{t_1}^B.\eqno(8)$$
 So we can get
$$
\begin{array}{lll}
& \ \  &P(\tau^B>t_1,\tau^C>t_2)\\
&=&E^B[I_{(\tau^B> t_2)}\exp\{-\int^{t_1}_0 (b_0+I_{(\tau^C\leq t)}\frac{-b_1}{(t-\tau^C+1)^{b_2}}dt)\}]\\
&=&E^B\left\{I_{(\tau^C> t_2)}e^{-b_0t_1}
    \exp\{-b_1I_{(\tau^C\leq t_1)}-\int^{t_1}_{\tau_C} \frac{1}{(t-\tau^C+1)^{b_2}}dt\}\right\}
 \end{array}
$$
If $b_2\neq 1$, the above equation equals to
$$
\begin{array}{lll}
&=&e^{-b_0t_1}E^B[I_{(t_2<\tau^C\leq t_1)}\exp\{\frac{-b_1}{1-b_2}[(t_1-\tau^C+1)^{1-b_2}-1]+I_{(\tau^C> t_1)}]\\
&=&\exp\{-b_0t_1-\frac{b_1}{1-b_2}\}[\int^{t_1}_{t_2}c_0e^{-c_0t}[b(t_1-t)+1]bt+e^{-c_0t_1}]\\
&=&b(t_1-t_2+\frac{1}{b}-\frac{1}{c_0}){e^{-c_0t_2-b_0t_1}}+\frac{b}{c_0}e^{-(c_0+b_0)t_1}\\
&=&E^B\left\{I_{(\tau^C> t_2)}e^{-b_0t_1}\exp\{I_{(\tau^C\leq t_1)}\ln[b(t_1-\tau^C)+1]\}\right\}\\
&=&e^{-b_0t_1}E^B[I_{(t_2<\tau^C\leq t_1)}(b(t_1-\tau^C)+1)+I_{(\tau^C> t_1)}]\\
&=&e^{-b_0t_1}[\int^{t_1}_{t_2}c_0e^{-c_0t}[b(t_1-t)+1]bt+e^{-c_0t_1}]\\
&=&b(t_1-t_2+\frac{1}{b}-\frac{1}{c_0}){e^{-c_0t_2-b_0t_1}}+\frac{b}{c_0}e^{-(c_0+b_0)t_1},\\
 \end{array}
 \eqno{(9)}
$$
where $E^C$ denotes the expectation under the measure $P^C$. The
first equation follows from the definition of $P^C$, and the
fourth from the fact that $\lambda_t^B=b_0$ for $t<t_2$ under
$P^C$. By a similar argument for  $t_2\leq t_1\leq T $, the joint
distribution is given by
$$P(\tau^B>t_1,\tau^C>t_2)=b(t_1-t_2+\frac{1}{b}-\frac{1}{c_0}){e^{-b_0t_1-c_0t_2}}
+\frac{b}{c_0}e^{-(b_0+c_0)t_1}.\eqno(10)$$ The differentiation of
$P(\tau^B>t_1,\tau^C>t_2)$ with respect to $t_1$ and $t_2$ gives
the joint density of the default times
$$ f(t_1,t_2)=\frac{\partial^2P(\tau^B>t_1,\tau^C>t_2)}{\partial t_1\partial t_2}=\left\{
\begin{array}{ll}
c{b_0c_0}[t_2-t_1+\frac{1}{c}-\frac{1}{c_0}]e^{-b_0t_1-c_0t_2},&{\mbox {for}}\   t_1< t_2\leq T  \\
b{b_0c_0}[t_1-t_2+\frac{1}{b}-\frac{1}{b_0}]e^{-b_0t_1-c_0t_2},&
{\mbox {for}}\ t_2<t_1\leq T ,
\end{array}
\right.\eqno(11)
$$
\par The proof is completed.$\sharp$\par
{\bf Remark 1}\ \ It is worth noting that if $c{b_0}\neq b{c_0}$,
then $f(t_1,t_2)$ is not continuous on the plane $t_1=t_2$.
\par
{\bf Corollary 1}\ \ Under the assumption of Proposition 1, if
$b_1=c_1=0$, then the joint distribution of default times
$(\tau^B,\tau^C)$ with the default intensity defined by Eq.(3)(4)
on $[0,T]\times [0,T]$ is
$$P(\tau^B>t_1,\tau^C>t_2)={e^{-b_0t_1-c_0t_2}},\eqno(12)$$
in other words, when $b_1=c_1=0$, $(\tau^B, \tau^C)$ are
default-independent on $[0,T]\times [0,T]$.
\par
{\bf Proof}\ \ We can get (11) by taking limit in Eq.(9) or Eq.
(10) as $c\rightarrow 0^+$. $\sharp$\par\bigskip
\par
{\bf Corollary 2}\ \ If $-b_1=b_2=b>0,-c_1=c_2=c>0$, the marginal
distributions of $\tau^B,\ \tau^C$ on $[0,T]$ are given by
$$  P(\tau^B>t_1)={e^{-b_0t_1}}+\frac{b}{c_0}e^{-b_0t_1}[e^{-c_0t_1}-1+{c_0}t_1],\ t_1\leq T \eqno(13)$$
$$P(\tau^C>t_2)={e^{-c_0t_2}}+\frac{c}{b_0}e^{-c_0t_2}[e^{-b_0t_2}-1+b_0t_2],\ t_2\leq T .\eqno(14)$$
\par{\bf Proof}\ \
We can get Eqs.(13) and (14) by taking $t_1=0$ and $ t_2=0$ in
Eqs.(9) and (10) respectively. $\sharp$
\par
{\bf Remark 2}\ \ The first term in Eq.(13) denotes the
firm-specific survival probability, and the second one denotes the
increment of firm B's survival probability because of the default of
firm C and the geometric  attenuation speed. As the result of
$$e^{-c_0t_1}-1+{c_0}t_1\leq \frac{1}{2}c_0^2t_1^2,$$
 the increment of firm B's survival probability satisfies
$$\frac{b}{c_0}e^{-b_0t_1}[e^{-c_0t_1}-1+{c_0}t_1]\leq \frac{1}{2}bc_0t_1^2e^{-b_0t_1}
\leq \left\{
\begin{array}{ll}
\frac{2c_0}{b_0}e^{-2}\ \ &{\mbox {for}}\   \frac{2}{b_0}\leq T , \\
\frac{1}{2}bc_0T^2e^{-b_0T}\ \ & {\mbox {for}} \ \frac{2}{b_0}> T.
\end{array}
\right.$$ From the above inequation we can get that the increment
of firm B's survival probability at time $t$ will be no more than
$\frac{1}{2}bc_0t_1^2e^{-b_0t_1}$. An analogous argument also
holds for firm C.
\par\bigskip
\begin{center}{\bf Section 3\ \ CDS Valuation in the Model of Dependent
Default\\ with Geometric  Attenuation function}
\end{center}
\par\bigskip
In this section we use the conclusion of Section 2 to price the
premium of  a CDS. A CDS is a contract agreement between
protection buyer and seller, in which the protection buyer pays
periodically to the protection seller a fixed amount fee(swap
premium pr spread) asking for a payment when the reference asset
defaults. A institute can use a CDS to transfer,  elude and hedge
the credit risk of a risky asset(or basket of risky assets) from
one party to the other. So a CDS is a very important instrument to
manage credit risk.
\par
Suppose interest rate $r_t$ is a constant $r$. Assume that party A
holds a corporate bond and faces the credit risk arising from
default of the bond issuer (reference party C). To seek protection
against such default risk, party A enters a CDS contract in which
he agrees to make a stream of periodic premium payments, known as
the swap premium to party B (CDS protection seller). In exchange,
party B promises to compensate A (CDS protection buyer) for its
loss in the event of default of the bond (reference asset).
Without loss of generality, we take the notional to be \$ 1 and
assume zero recovery under default. Firm B pays firm A after a
settlement period $\delta$ when the reference asset c defaults.
 Furthermore in Leung and Kwok (2005), they conclude the expression for the swap premium has little
  dependence on the default intensity of the protection buyer, so we impose that during the entire contract,
  firm A doesn't default. \par
  The default intensity processes of firm B and C are given by
  Eqs.(3) and (4) in special cases
 $$\lambda_t^B=b_0-I_{(\tau^C\leq t)}\frac{b}{b(t-\tau^C)+1}\ ,\eqno{(3)}$$
$$\lambda_t^C=c_0-I_{(\tau^B\leq t)}\frac{c}{c(t-\tau^B)+1}\ .\eqno{(4)}$$
Since it takes no cost to enter a CDS, the value of the swap premium
$S(T)$ is determined by
$$\begin{array}{lll}
&
&\sum_{i=1}^nE[e^{-rT_i}S(T)I_{(\tau^B\bigwedge\tau^C>T_i)}]+S(T)A(T)\\
&=&E[e^{-r(\tau^C+\delta)}I_{(\tau^C\leq
T)}I_{(\tau^B>\tau^+\delta)}]
 \end{array}
\eqno(15)$$ where
$\{T_1,\ldots,T_n\}$ are the swap payment dates with
 $0=T_0<T_1<\ldots<T_n=T,\ T_i-T_{i-1}=\Delta T,\ T+\delta<T^*$,\
 and
$\delta $ is the length of settlement period. Here $\tau^C+\delta$
represents the settlement date at the end of the settlement
period. The first term in Eq.(15) gives the present value of the
sum of periodic swap payments(determined when either B or C
defaults or at maturity) and $S(T)A(T)$ is the present value of
the accrued swap premium for the fraction of period between
$\tau^C$  and the last payment date. The present value of accrued
swap premium is given by
$$S(T)A(T)=S(T)\sum_{i=1}^nE[e^{-r\tau^C}\frac{\tau^C-T_{i-1}}{\Delta T}
I_{(T_{i-1}<\tau^C\leq T_i)}I_{(\tau^B>\tau^C)}].\eqno(16)$$
\par
In the following, we will calculate all the expectations in
Eq.(16). For simplicity we denote $\beta:=b_0+c_0+r$.
\par
It is easy to get from Eq.(6)
$$E[I_{(\tau^B\bigwedge\tau^C>T_i)}]=P(\tau^B>T_i,\tau^C>T_i)=e^{-(b_0+c_0)T_i},\eqno(17)$$
So
$$\sum_{i=1}^nE[e^{-rT_i}S(T)I_{(\tau^B\bigwedge\tau^C>T_i)}]=S(T)
\frac{e^{-\beta\Delta T }(1-e^{-\beta T })}{1-e^{-\beta\Delta
T}}.\eqno(18)$$
 From Eq.(11) it can be found
$$\begin{array}{lll}
& &E[e^{-r(\tau^C+\delta)}I_{(\tau^C\leq T)}I_{(\tau^B>\tau^C+\delta)}]\\
&=&\int^T_0\int^{+\infty}_{t_2+\delta}e^{-r(t_2+\delta)}
b{b_0}{c_0}[{t_1-t_2+}\frac{1}{b}-\frac{1}{b_0}]
e^{-b_0t_1-c_0t_2}  dt_1dt_2\\
&=&\frac{bc_0}{\beta}{(\frac{1}{b}+\delta)}e^{-(r+b_0)\delta}[1-e^{-\beta
T}]
\end{array}
\eqno(19)
$$
and
$$
\begin{array}{lll}
& &E[e^{-r\tau^C}\frac{\tau^C-T_{i-1}}{\Delta
T}I_{(T_{i-1}<\tau^C\leq T_i)}I_{(\tau^B>\tau^C)}]
\\
&=&\int^{T_i}_{T_{i-1}}\int^{+\infty}_{t_2}e^{-rt_2}\frac{t_2-T_{i-1}}{\Delta
T}b{b_0}{c_0}[{t_1-t_2}+\frac{1}{b}-\frac{1}{b_0}]e^{-b_0t_1-c_0t_2}dt_1dt_2\\
&=&\frac{c_0}{{\beta}\Delta T}[T_{i-1}e^{-\beta
T_{i-1}}-T_{i}e^{-\beta T_{i}}+
   (T_{i-1}-\frac{1}{\beta})(e^{-\beta T_{i}}-e^{-\beta
   T_{i-1}})].
\end{array}\eqno(20)
$$
 So
$$\begin{array}{lll}
S(T)A(T)
&=&S(T)\frac{c_0}{\beta\Delta T}[\sum_{i=1}^n(T_{i-1}e^{\beta T_{i-1}}-T_{i}e^{\beta T_{i}})\\
& &+\sum_{i=1}^nT_{i-1}(e^{-\beta T_{i}}-e^{\beta
T_{i}})-\frac{1}{\beta}\sum_{i=1}^n(e^{-\beta T_{i}}-
e^{-\beta T_{i-1}})]\\
&=&S(T)\frac{c_0}{\beta\Delta T}\{-Te^{-\beta T}-\frac{1}{\beta}(e^{-\beta T}-1)\\
& &+\frac{1}{(1-e^{-\beta \Delta T})}[{Te^{-\beta T }- \Delta
Te^{-\beta \Delta T})-(T-\Delta T)e^{-\beta(T+ \Delta T)}}]
\}\\
&=&S(T)\frac{c_0}{\beta^2\Delta T}\frac{1}{(1-e^{-\beta \Delta
T})}[1-e^{-\beta \Delta T}-\beta\Delta T e^{-\beta \Delta T}].
\end{array}
$$
That is to say $$ A(T)=\frac{c_0}{\beta^2\Delta
T}\frac{1}{(1-e^{-\beta \Delta T})}[1-e^{-\beta \Delta
T}-\beta\Delta T e^{-\beta \Delta T}].\eqno(21)$$
Take Eqs.(18),(19)
and (21) into (15), we can get
\par
{\bf Proposition 2}\ \ Assume the default buyer doesn't default
during the entire contract, and the default intensities of B (the
protection seller) and C(protection buyer) are given by Eq.(3) and
(4), then the swap premium is given by
$$S(T)=\frac{bc_0}{\beta}{(\frac{1}{b}+\delta)}e^{-(r+b_0)\delta}[1-e^{-\beta T}]
\times[\frac{e^{-\beta\Delta T }(1-e^{-\beta T })}
{1-e^{-\beta\Delta T }}+A(T)]^{-1}. \eqno(22)$$ where $A(T)$ is
given by Eq.(21).
\par\bigskip
{\bf Remark 2:}\ \ Due to
$$1-e^{-\beta \Delta T}-\beta\Delta T e^{-\beta \Delta T}\geq \frac{1}{2}\beta^2\Delta T^2e^{-\beta\Delta T},$$
 the swap premium $S(T)$ is bounded by
$$
S(T)\leq\frac{bc_0}{\beta}{(\frac{1}{b}+\delta)}e^{-(r+b_0)\delta}
\frac{(1-e^{-\beta T})(e^{\beta\Delta T}-1)}{\frac{c_0\Delta
T}{2}+1-e^{-\beta T}}$$

{\centerline {\bf Section 4. Conclusion}}
\par\bigskip
In this paper, a geometric   function is introduced to reflect the
attenuation speed of impact of one firm's default to its partner. If
the two firms are competitions (copartners),
 the default intensity of one firm will decrease (increase) abruptly when the other firm defaults.
 As time goes on, the impulsion will decrease gradually until extinct. In this model,
 the joint distribution and marginal distributions of default times are derived by employing
the change of measure,  so can we value the fair swap premium of a
CDS and get the analytic expression.
\par
\bigskip
{\bf Reference}\par\bigskip
\def\hang{\hangindent\parindent}
 \def\textindent#1{\indent\llap{#1\enspace}\ignorespaces}
 \def\re{\par\hang\textindent}
\re{[1]} Collin-Dufresne P, Goldstein R S, Hugonnier J. A general
formula for valuing defaultable securities. $Econometrica$,
2004,72:1377-1407. \re{[2]} Davis M, Lo V. Infectious defaults.
$Quantitive\ Finance$,2001,1:383-387. \re{[3]}Duffie, D., and K.
Singleton (1999): "Modeling Term Structures of Defaultable Bonds,"
Review of Financial Studies, 12:687-720.
 \re{[4]}Harrison M, Pliska S. Martingales and stochastic
integrals in the theorey of continuous trading. $Stochastics\
processes\ and\ their\ applications$,1981, 11:215-260.
\re{[5]}Jarrow R A, Yu F. Counterparty risk and the pricing of
defaultable securities. $Journal$ $of \ Finance$, 2001,56(5):
1765-1799. \re{[6]}Kijima M,Muromachi Y. Credit Events and the
Valuation of Credit Derivatives
 of Basket Type. $Review\ of\ Derivatives\ Research$, 2000,4:55-79.
\re{[7]}Lando D.Three essays on contigent claims pricing,
Ph.D.dissertation,Cornell University,1994.
 \re{[8]}Lando D. On Cox processes and credit risky securities.$ Review\ of\ Derivatives\
Research $, 1998,2:99-120.
 \re{[9]}Laurent J-P Gregory J.
Basket Default Swaps, CDO's and Factor Copulas,
 $Working Paper$, ISFA Actuarial School, University of Lyon,2003.
\re{[10]}Leung S Y,Kwok Y K. Credit default swap valuation with
counterparty risk. $Kyoto\ Economic$\  $Review$,2005,74(1):25\
-45.
 \re{[11]} Li D X. On Default Correlation: A Copula
function Approach. $ Journal\ of\ Fixed\ Income$, 2000,Vol 9,
March:43-54. \re{[12]} Schonbucher P, Schubert D. Copula-dependent
default risk in intensity models. $Working\ paper$, Bonn
University, 2001.
 \re{[13]} Shreve S E. Stochastic
Calculus for Finance I:The Binomial Asset Pricing Model. Springer,
New York. 2004. \re{[14]} Zhou C. An analysis of default
correlations and multiple defaults. $The\ Review\ of \ Financial\
Studies$,  2001,14(2):555-576.
\end{document}